\documentclass[useAMS,usenatbib]{mn2e}

\usepackage{graphicx}
\def\hmpc{\ {\rm {\it h}^{-1}Mpc}}
\def\etal{{\it et al. }}

\title[Monthly Notices: \LaTeXe\ guide for authors]
  {The Redshift-Space Two Point Correlation Function of ELAIS-S1 Galaxies }
\author[V. D'Elia et al.]
  {V.~D'Elia,$^1$
  E.~Branchini,$^2$ F.~La Franca,$^2$ V.~Baccetti,$^2$
  I.~Matute,$^3$ 
  \newauthor 
  F.~Pozzi,$^4$ C. Gruppioni$^5$\\
  $^1$INAF-Osservatorio Astronomico di Roma
     via Frascati 33, Monteporzio-Catone (RM), I-00040 Italy\\
  $^2$Dipartimento di Fisica, Universit\'a degli Studi ``Roma Tre'',
      Via della Vasca Navale 84, I-00146, Roma, Italy\\
  $^3$Max-Planck Institut f\"ur extraterrestrische Physik (MPE),
      Giessenbachstrasse, Postfach 1312, 85741 Garching, Germany\\
  $^4$Dipartimento di Astronomia, Universit\`a di Bologna,
      viale Berti Pichat 6, I-40127 Bologna, Italy\\
  $^5$INAF-Osservatorio Astronomico di Bologna, 
      via Ranzani 1,I-40127 Bologna, Italy}
\date{Released 2005 January 14}

\pagerange{\pageref{firstpage}--\pageref{lastpage}} \pubyear{2002}

\def\LaTeX{L\kern-.36em\raise.3ex\hbox{a}\kern-.15em
    T\kern-.1667em\lower.7ex\hbox{E}\kern-.125emX}

\begin{document}

\label{firstpage}

\maketitle

\begin{abstract}
We investigate the clustering properties of galaxies in the recently
completed ELAIS-S1 redshift survey  through their spatial two point
autocorrelation function.  We used a sub-sample of the ELAIS-S1
catalog covering approximately $4$ deg$^2$ and consisting of 148
objects selected at $15 \mu m$ with a flux $>0.5$ mJy and redshift
$z<0.5$.  We detected a positive signal in the correlation function
that, in the range of separations $1-10 \hmpc$ is well approximated by a power law
with a slope $\gamma=1.4\pm0.25$ and a correlation length $s_0=5.4
\pm1.2 \hmpc$, at the 90 \% significance level.  This result is in
good agreement with the redshift-space correlation function measured
in more local samples of mid infrared selected galaxies like the IRAS
PSC$z$ redshift survey.  This suggests a lack of significant
clustering evolution of infrared selected objects out to $z=0.5$ that
is further confirmed by the consistency found between the correlation
functions measured in a local ($z<0.2$) and a distant ($0.2<z<0.5$)
subsample of ELAIS-S1 galaxies. We also confirm that optically selected galaxies
in the local redshift surveys, especially those of the SDSS sample,
are significantly more clustered than infrared objects.
\end{abstract}


\section{Introduction}

Investigating the redshift-space distribution of galaxies has long been 
regarded as a fundamental aspect of observational cosmology.
The primary statistical tool for characterizing galaxy clustering is
the spatial two point correlation function, $\xi(s)$
since, in the current paradigm
of structure formation, the galaxy two-point correlation function
is directly related to the initial power spectrum of mass density fluctuations.

This is true as long as galaxies trace the underlying mass density
field.
However, it is now well established that the clustering of galaxies
at low redshift depends on a variety of factors, implying that 
not all types of galaxies can be regarded as unbiased mass tracers.
The clustering of optically selected galaxies has been found to depend
on galaxy luminosity (Norberg \etal 2002 and reference therein),
morphological and spectral type (Hermit \etal 1996, Zehavi \etal 2002,
Magdwick \etal 2003). On the other hand, the clustering of infrared selected 
galaxies seems to depend on their infrared color (Hawkins \etal 2001)
rather than  on luminosity (Szapudi \etal 2000).

The evidence that different galaxy populations might give different
biased pictures of the mass distribution has complicated but also 
enriched the interpretation of galaxy clustering. Indeed, the very 
fact that the spatial clustering of galaxies
is related to their physical properties represents an
important observational test for all theories of galaxy formation.
In particular, strong constraints on galaxy evolution models 
can be obtained by measuring the relative clustering of different 
extragalactic objects as a function of redshifts, for which very deep
galaxy samples are required.

Several deep redshift surveys of optically selected galaxies, like the 
Keck surveys of the GOODS-north (Wirth \etal 2004) and GOODS-south
(Vanzella \etal 2004) fields and the DEEP2 redshift survey (Davis \etal
2003), are currently being performed. First results 
based on early data look very promising indeed.
The analysis of Coil \etal (2004)  has shown that the two-point correlation
function of DEEP2 galaxies with a median redshift $z=1.14$ is consistent
with that measured by Adelberger\etal (2003) in a very deep ($z \sim 3$)
sample of Lyman break galaxies but is significantly smaller than the correlation
measured in the local ($z\sim0$) 2dF galaxy sample.
Since 2dF $b_j$-selected galaxies are known to trace the underlying mass 
density field in a unbiased way
(Verde \etal 2002, Lahav \etal 2002), the smaller correlation measured
in  DEEP2 survey implies that this is not a strongly biased sample of 
objects either, since the clustering of the dark mass is expected to
decrease with redshift in a similar way (Coil \etal 2004).

Mid-infrared selected galaxies also constitutes a very interesting 
population of objects since they are also known to trace the 
underlying mass distribution in the local universe in an almost unbiased fashion.
More precisely, it has been found that 
the  fluctuations in their number density, $\delta_g$ is related to the 
underlying mass overdensity field $\delta_m$ through a simple, linear biasing
relation $\delta_g \sim 1.2 \delta_m$ (Tegmark, Zaldarriaga \& Hamilton 2001
and Taylor \etal 2002). 
Another reason why mid-infrared selection is
interesting is that luminosity in this band is approximately proportional to
star-formation rate (independent of dust), thus a mid-infrared sample of  
galaxies will highlight the distribution of star-formation activity at 
a particular epoch.
Clearly, it would be very important to 
quantify the clustering evolution of mid-infrared
selected objects and compare it with that of optically selected galaxies.
This is indeed the  main goal of this work
in which we analyze the clustering properties of sample of 
mid-infrared selected galaxies extracted from the ELAIS redshift survey 
(Oliver \etal 2000) and extending out to a redshift $z=0.5$.
According to La Franca \etal (2004), two main spectroscopic classes
have been found to dominate the extragalactic population of these
objects: star-forming galaxies (from absorbed to extreme starbursts:
$\nu L\nu (15\mu m)\sim 10^8-10^{11} L\odot$), which account for 75\%
of the sources, and active galactic nuclei (excluded from this
analysis) which account for 25\% of the sources.  About 20\% of the
extragalactic ELAIS sources are dust-enshrouded starburst
galaxies, while passive galaxies are essentially absent from the sample.

Our analysis is performed in redshift-space and thus complements the 
previous work of Gonzalez-Solares \etal. (2004) who measured the angular 
correlation properties of a similar sample of ELAIS galaxies that 
from which they have inferred their spatial correlation properties.

The outline of this paper is as follows. In section 2
we describe the ELAIS-S1 galaxy redshift survey that we analyze in this work. 
In section 3 we discuss our method of estimating the 
two-point correlation function, assess its robustness and evaluate 
its statistical uncertainties. The main results are presented in 
section 4,
and discussed in section 5, 
in which we also draw our main conclusions.

\section{The ELAIS S1 Sample}

The European Large-Area {\it ISO} survey (ELAIS, Oliver \etal 2000;
Rowan-Robinson \etal 2004) is the largest Open Time programme
conducted by the {\it ISO} satellite (Kessler \etal 1996). It covers
an area of $12$ deg$^2$, divided in four fields (N1, N2 and N3 in the
northern hemisphere and S1 in the southern one ) distributed across
the sky in order to decrease the biases due to cosmic variance. The
survey bands are at $6.7$, $15$, $90$ and $170 \mu m$; the $15 \mu m$
one presents the highest density of galaxies (Serjeant
\etal 2000, Gruppioni \etal 2002, 
La Franca \etal 2004), making it the best choice for a study of their
clustering properties.  In this work we concentrate on the southern
area, S1.

This survey is made of nine raster observations, each covering $40
\times 40 $ arcmin$^2$.  The final analysis catalog at 15 $\mu$m in
the S1 field has been released by Lari \etal (2001) covering an area
of $2\times2$ deg$^2$ centered at $\alpha (2000) = 00^h$ $34^m$
$44.4^s$, $\delta (2000) = -43^{\circ}$ $28'$ $12''$. It includes 462
mid-IR sources down to a flux limit of 0.5 mJy.  We have restricted
our analysis to a highly reliable subsample of 406 objects (La Franca \etal 2004).

A detailed description of the optical classification of the ELAIS-S1
sources, size and completeness function of the areas used in our
study, as well as the observed counts for each class of sources
(normal galaxies, Starburst galaxies and AGN) are presented by La
Franca \etal (2004). The measure of the evolution of star-forming
galaxies has been investigated by Pozzi \etal (2004)
and Gruppioni \etal (2005), while a first estimate
of the luminosity function for type-1 AGN has been presented by Matute
\etal (2002)
\footnote{Data and related papers about the ELAIS southern survey are
available at: {\tt http://www.fis.uniroma3.it/ $\sim$ ELAIS$_{\rm}$ }}.

The central raster, S$1\_{\rm R}5$, has been observed three times and thus
represents the deepest part of the sample.  As a consequence the
ELAIS-S1 area has been divided into two regions: a) the central and
deepest part (S$1\_{\rm R}5$) which covers 0.55 deg$^2$ and reaches a 20\%
completeness at fluxes of 0.7 mJy, and b) the remaining area (S1$\_{\rm
Rest}$) which covers 3.65 deg$^2$ and reaches a 20\% completeness at
fluxes of 1.1 mJy.

The optical analysis of the $406$ objects brings $332$ (80\%) optical
identification on CCD exposures down to R$\sim$23, and $290$
spectroscopic classifications (90\% of the optically identified
sample).  Of these, $93$ were found to be stars and $199$
extragalactic sources. Among these, we kept only the galaxies with
$z<0.5$, because according to analyses based on the mean
optical/mid-IR ratios of the sources, the sample is virtually $100\% $
spectroscopically complete down to this redshift (La Franca
\etal 2004). We have excluded all the AGN type sources. 
This left us with $148$ redshift determined 
galaxies, $48$ belonging to the deeper S$1\_{\rm R}5$ raster and the
remaining $100$ galaxies to the surrounding areas S1$\_{\rm rest}$.

The angular distribution of ELAIS-S1 galaxies in our sample is displayed in 
Fig.1 (filled dots). The galaxy surface 
density is higher in the central region
corresponding to the deeper S$1\_{\rm R}5$ raster.
\begin{figure}
\centering
\includegraphics[angle=0,width=8cm]{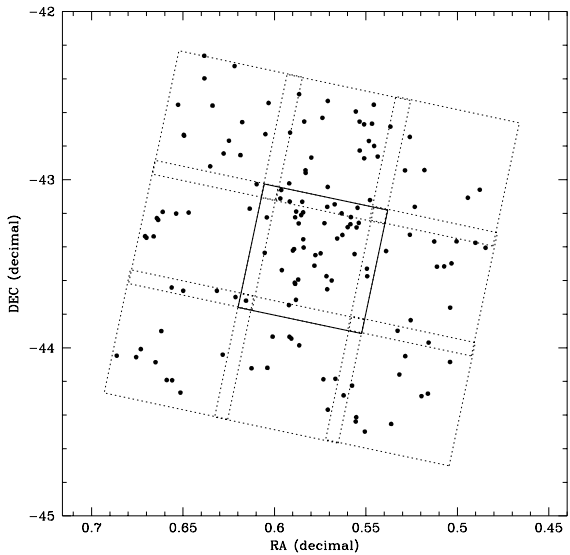}
\caption{The angular distribution of ELAIS-S1 sources. The solid square 
is the central raster S$1\_{\rm R}5$, observed three times.} 
\label{2d}
\end{figure}

The continuous line histograms in both panels of
Fig. 2 shows the redshift distribution of the ELAIS-S1 galaxies 
in the S$1\_{\rm R}5$ (top) and S1$\_{\rm Rest}$ (bottom) rasters. 
Gonzalez-Solares \etal. (2004) 
have shown that the redshift distribution of ELAIS objects obtained from
follow-up spectroscopic observations and photometric redshifts is consistent with that predicted
from  the ELAIS luminosity function of Pozzi \etal (2004). 
This redshift distribution will used in the 
next section to estimate the galaxy correlation function and to
construct the mock ELAIS catalogs.

\begin{figure}
\centering
\includegraphics[angle=0,width=8cm]{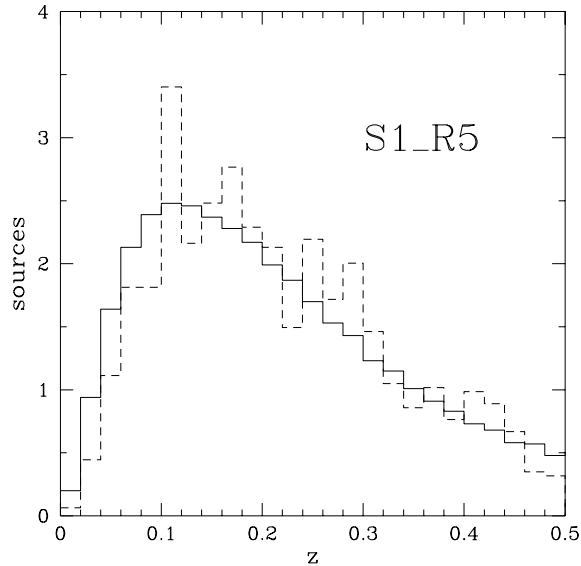}
\includegraphics[angle=0,width=8cm]{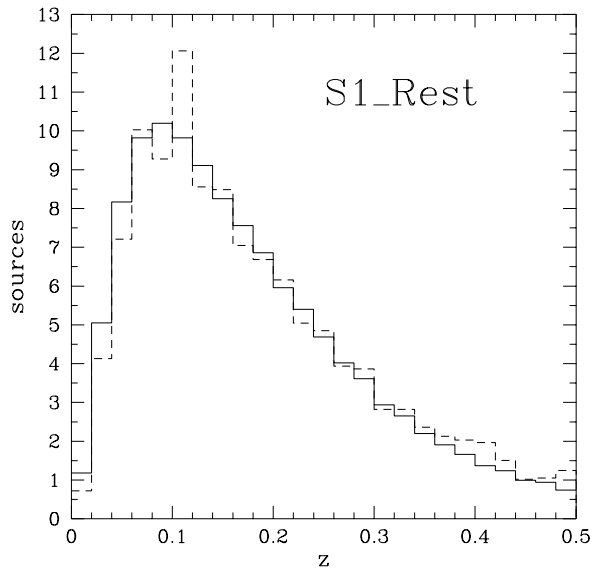}
\caption{The redshift distribution of the sources in the S$1\_{\rm R}5$ raster (top panel)
and in S1$\_{\rm Rest}$ (bottom panel). 
The solid lines histogram indicates the redshift distribution of galaxies in the real ELAIS-S1
sample, while the dashed-line histogram indicates the average redshift distribution
of objects in the 30 mock ELAIS-S1 catalogs.
}
\label{nz}
\end{figure}

\section{Measuring the redshift space correlation function $\xi(s)$}

Our purpose is to estimate the redshift-space two point correlation
function of ELAIS-S1 galaxies and its uncertainties.  It is worth
stressing here that clustering in redshift-space is systematically
different from the true one in real-space. On small scale, velocity
dispersion in virialized systems smears out structure along the line
of sight, hence depressing the amplitude of $\xi(s)$ at small
separation. On large scales, coherent infall motions in high density
regions or outflows from low density environments enhance structure
along the line of sight and thus increase the amplitude of $\xi(s)$ on
the corresponding scales.  Here we make no attempt of correcting for
redshift-space distortions. Therefore, direct comparisons will only be 
made with similar measurements of the redshift-space
two point correlation function of other galaxy redshift catalogs.

\subsection{Estimate of $\xi(s)$}

Several different methods have been proposed to
estimate the two point correlation function. In this work we use
the Landy \& Szalay (1993) [LS] estimator that, according to extensive numerical
tests, turned out to be the best suited for astrophysical applications 
(Kerscher, Szapudi and Szalay 2000).

With the LS estimator a reliable and unbiased estimate of $\xi(s)$ is obtained by
cross-correlating the real dataset with a sample of fake objects
lacking any spatial correlation which, however, suffers from the same selection 
effects and incompleteness of the real catalog.
To obtain a sample of synthetic objects we first create a random sample of points
uniformly distributed over the same area as the real dataset, and then
we generate their redshift distribution constrained to match that of the real objects
determined by Pozzi \etal (2004).
For each random catalog we generate 10 times the number of fake 
objects as real galaxies.

The LS estimator is:
\begin{equation}
\xi(s) =  
{{N_{DD}(s)}\over{N_{RR}(s)}} \left( {n_R}\over{n_D} \right)^2 
-2 {{N_{DR}(s)}\over{N_{RR}(s)}} \left({n_R}\over{n_D}\right)
+ 1,
\label{csi}
\end{equation}
where $N_{DD}, N_{DR}$ and $N_{RR}$ as the weighted data-data, data-random and random-random 
pair counts with redshift-space separation $s$. $n_D$ and $n_R$ are the 
mean densities of galaxies and random objects, respectively.

The flux limited nature of our dataset is accounted for by weighting each galaxy
by 
\begin{equation}
w_i =   {{1}\over{1+4\pi n(z_i)J_3(s)}},
\label{weight}
\end{equation}
where $n(z_i)$ is the observed space density of galaxies  at 
redshift $z_i$ and $J_3(s)=\int_0^s\xi(r)r^2dr$ (Efstathiou 1988).
To implement this weighting scheme, that requires prior knowledge of $\xi(s)$,
we have adopted an iterative procedure in which we first compute
$\xi(s)$ using $w_i=1$ and then we fit a power law to the results in the range
$1-10 \hmpc$. Then we compute the new weights $w_i$ by truncating the maximum
value of $J_3$ to 1500. Only one iteration was needed to obtain a stable result.

To check the robustness of our results we have varied both the estimator
of $\xi(s)$  and the weighting scheme. More precisely, we  have also used the estimators  
of Hewett (1982), Davis \& Peebles (1983) and Hamilton (1993). Moreover, for the case of the 
LS estimator only, we have 
implemented two alternative weighting schemes: the case of no weight (i.e. $w_i=1$) and 
the case in which we weight by 
the inverse of the  selection function (i.e. $w_i=n_D/n(z_i)$).
In all cases explored we have found  that the various estimates of $\xi(s)$ 
were consistent, within the errors, in the range $1-10 \hmpc$.

\subsection{Errors estimation}

To quantify the uncertainties in the estimate of $\xi(s)$
we have measured the correlation function in a sets of 30
independent mock galaxy catalogs designed to mimic the
properties of the real ELAIS-S1 sample.

These mock catalogs
were constructed from the N-body numerical experiment 
labelled $\mathrm{L3S}$ performed by Cole \etal (1998).
Their simulation assumes a flat $\Lambda$CDM model
cosmology with $\Omega_m=0.3$, $\Lambda=0.7$
an rms fluctuation of the mass contained in spheres of radius $8\hmpc$,
$\sigma_8=1.13$ and 
and a CDM power spectrum  with shape parameter $\Gamma=0.25$. 
The simulation box side is $345.6\hmpc$ and has $192^3$
particles. 

To construct each mock catalog we have used three different output 
of the simulations (corresponding to redshifts z=0, z=0.33 and z=0.5)
and  took advantage of the
periodic boundary conditions 
to obtain different configurations of points in each box
by re-centering the coordinate system on a
particle chosen at random.
We have then stacked six boxes
in order of increasing redshift (three of them at z=0,
two at z=0.33 and one at z=0.5), placed the observer
at the center of the first box and identified the
X coordinate with the direction of stacking.
We have then identified the direction of the X axis with
the center of the ELAIS field and considered all particles
within an area of $2\times2$ deg$^2$ from the 
field centre.
To assign each particle a redshift we have used the 
distance to redshift relation for a flat, $\Lambda$CDM
universe
\begin{equation}
r=
{{c}\over{H_o}}
\int^z_0 {{dz\prime}\over{\sqrt{\Omega_m(1+z\prime )^3+\Omega_{\Lambda}}}},
\label{zr}
\end{equation}
where $r$ is the comoving distance, and then added the 
line of sight component of the particle peculiar velocity.
Finally, a population of mock galaxies have been 
extracted from the particles through a
Montecarlo rejection procedure designed 
to match the observed redshift distribution of real
ELAIS galaxies in both the inner
$40 \times 40 $ arcmin$^2$ S$1\_{\rm R}5$-like, and the outer  
S1$\_{\rm Rest}$-like samples.
This procedure has been repeated to obtain 30 mock ELAIS-S1
samples,  containing $\sim 120$ objects each, for which we
have evaluated the two-point correlation function.
The mean redshift distribution of fake objects in the 30
catalogs is shown in of Fig. 2 (dashed line histogram)
for the  S$1\_{\rm R}5$ (top panel) and S1$\_{\rm Rest}$ (bottom) rasters

The filled dots in Fig. 3 shows the 
average $\xi(s)$ in the 30 mock catalogs. 
The errorbars  represent the rms scatter around the mean values.
They constitutes our estimate of errors in the 
measurement of two point correlation function of the
real ELAIS-S1 galaxies
These errors account for cosmic variance and sample noise,
that constitutes the main source of uncertainty. The 
large size of  the sample guarantees
that the 'integral constraint' correction 
is only ~5\% (Gonzalez-Solares \etal 2004)
and thus can be neglected in the error budget.

The open dots in Fig. 3 shows 
the ``true'' $\xi(s)$ of the simulation,
measured by considering all particles in one of our 
ELAIS-like samples. The ``true'' $\xi(s)$ 
is consistent with 
the average $\xi(s)$ of the mocks, 
indicating that our method for estimating $\xi(s)$ is indeed 
unbiased.

The mock  ELAIS-S1 catalogs mimic the geometry and
selection effects of the real sample and account for
the spatial clustering and its evolution in a flat $\Lambda$CDM universe.
However, they are not guaranteed to reproduce
the clustering properties of the ISO galaxies unless, of course, these
trace the underlying mass density field out to $z=0.5$. 

As we have verified {\it a posteriori} this is indeed the case,
in the sense that the ``true'' $\xi(s)$ 
in the range $1-10 \hmpc$  is well approximated by a power law 
whose best fit parameters, displayed in 
Fig. 3 are consistent within the errors with those
determined in the
 analysis of the real ELAIS-S1 sample presented in the 
following section.

\begin{figure}
\centering
\includegraphics[angle=0,width=8cm]{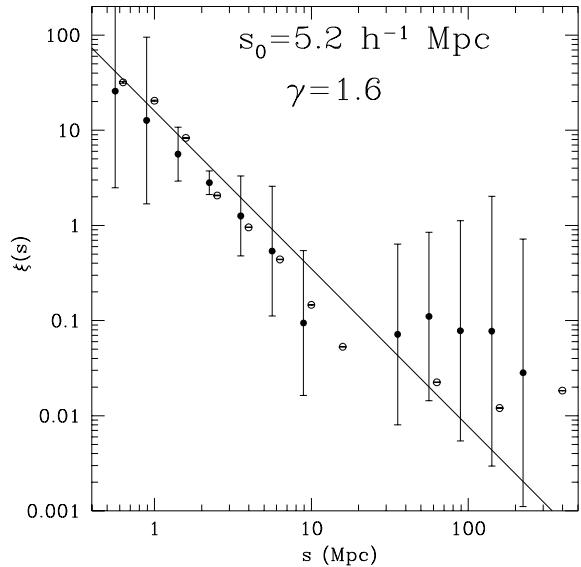}
\caption{Average $\xi(s)$ measured in the 30 mock ELAIS-S1 catalogs (filled dots)
and the rms scatter around the mean (errorbars). Open dots shows the 
``true'' $\xi(s)$ 
of all particles in the simulation contained in a region with the  
same volume of the ELAIS-S1 sample.
The straight line represents the best power law fit to the 
``true'' $\xi(s)$  in the range  $1-10 \hmpc$. The 
values of the best fit parameters are also shown. 
}
\label{spe1}
\end{figure}

\section[]{Results}
\label{sec:results}  
 
The filled dots plotted in Fig. 4 represent the two-point correlation
function of the ELAIS-S1 galaxies computed using the LS estimator.
The errorbars, evaluated from the 30 ELAIS-S1 mock catalogs, are the same shown in  Fig. 3.
Clearly, $\xi(s)$ is well approximated by a power law out to separations
of  $ 10 \hmpc$. In the range $1-10 \hmpc$ the
best fit power law model $\xi(s) = (s/ s_0)^{-\gamma}$,
has a correlation length of $s_0 = 5.4\pm 1.2 \hmpc$ 
and a slope $\gamma = 1.45\pm 0.25$, both determined at the $90 \%$
confidence level. Including the correlation of galaxy pairs
at separation $s=0.6 \hmpc$, also shown in the figure, does not 
modify this result appreciably.

Breaking down the sample by redshift does not change results significantly either.
Indeed, in the {\it local} sample composed by 82 objects at $z<0.2$ 
$\xi(s)$ is still well approximated by a power law in the range $1-10 \hmpc$ 
with best fit parameters
$s_0 = 5.4 \pm 1.6 \hmpc$ and  $\gamma = 1.6\pm 0.4$.
These values agree 
with those found for the {\it distant} sample of 66 objects at at $0.2 \le z<0.5$ 
for which we have found $s_0 = 5.1\pm 1.6 \hmpc$ and  $\gamma = 1.4\pm 0.4$,

Finally, we have verified that evaluating errors 
from 100 bootstrap realizations of the 
ELAIS-S1 sample rather than from the mock catalogss
does not change significantly the results as
the correlation length only decreases by $\sim 10 \%$ 
and the slope becomes $\sim 4 \%$ flatter.

\begin{figure}
\centering
\includegraphics[angle=0,width=8cm]{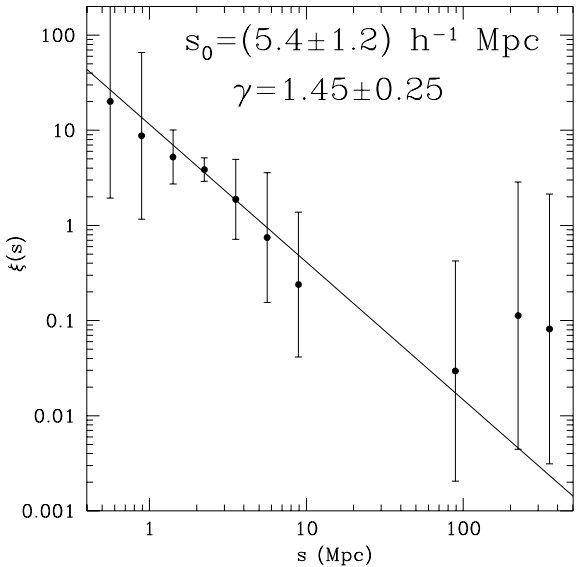}
\caption{The redshift-space correlation function for the ELAIS-S1 galaxy sample
(filled dots). Error bars have been calculated using the $30$ ELAIS-like mock catalogs.
The solid line represents the power law best fit $\xi(s) = (s/ s_0)^{-\gamma}$ 
to the data in the range $1-10 \hmpc$. The best fit parameters
are indicated in the plot.}
\label{csidata}
\end{figure}

\section[]{Discussion and Conclusions}
\label{sec:discuss}

\begin{table*}
  \centering
  \caption{Clustering of Different Galaxy Redshift Surveys: the Parameters of the Power Law Model}
  \begin{tabular}{lcccc}\hline
    \\
    Survey &  Selection Band & $\Delta$z & $s_0 (\hmpc)$ & $\gamma$ 
    \\
    \hline
    \\
    ELAIS-S1\footnote[1]{This work} & 15$\mu$m & 0.0-0.5 & $5.40 \pm 1.20$ & $1.45 \pm0.25$\\
    PSC$z$\footnote[2]{Hawkins \etal (2001)} & 60$\mu$m & 0.004-0.1 & $4.77 \pm 0.20$ & $1.30 \pm0.04$\\
    CfA2\footnote[3]{de Lapparent, Geller \& Huchra (1998)} &B-band &0.0-0.05 & $\sim 7.5$ & $\sim 1.6$ \\
    ORS\footnote[4]{Hermit \etal (1996)}&B-band &0.0-0.027 & $7.60 \pm 1.20$ & $1.60\pm 0.10$ \\
    LCRS\footnote[5]{Tucker \etal (1997)}&R-band &0.033-0.15   & $6.3 \pm 0.3$ & $1.86\pm 0.03$ \\
    SDSS\footnote[5]{Zehavi \etal (2002)}&$r^{\star}$-band &0.019-0.13 & $\sim 8.0$& $\sim 1.2$\\
    2dFGRS\footnote[6]{Hawkins \etal  (2003)}& ${\rm b}_j$-band&0.01-0.20 & $6.82\pm 0.28$ & $1.57\pm 0.07$ \\
    \\
    \hline 
    \end{tabular}    

    \centering
    \begin{tabular}{l}\hline 
    $^1$This analysis \\
    $^2$Hawkins \etal (2001) \\
    $^3$de Lapparent, Geller \& Huchra (1988) \\
    $^4$Hermit \etal (1996)  \\
    $^5$Tucker \etal (1997) \\
    $^6$Zehavi \etal (2002) \\
    $^7$Hawkins \etal  (2003) \\
    \hline
    \label{table:parameters}
  \end{tabular}
\end{table*}


In this work we have evaluated the redshift-space two point correlation
function of mid-infrared selected galaxies in the deep ($z \le 0.5 $)
 ELAIS-S1 catalog.
We found a significant, positive correlation signal at separations $\le 10 \hmpc$, where
the two point correlation function is well approximated by a power law model
$\xi(s) = (s/s_0)^{-\gamma}$ with a correlation length of $s_0 = 5.4{\pm 1.2} \hmpc$ 
and a slope $\gamma = 1.45\pm 0.25$, with errorbars referring to a 90\% confidence level.
These results have been obtained using the LS estimator for $\xi(s)$  
and by  evaluating the errors from a set of 30 mock ELAIS-S1 catalogs.
These results are robust, in the sense that they do not change significantly when
varying the method of  estimating $\xi(s)$ (see section 3.1)
or when using different strategies to assess the errors (sections 3.2
and 4).

It is interesting to compare our results with those 
obtained from similar analysis of galaxy clustering in 
redshift-space.
Table 1  shows our result together with the corresponding 
results obtained from some of the major galaxy redshift surveys,
characterized by their passband/wavelength of selection (column 2)
and the redshift ranges they cover (column 3).
It is worth stressing that deviations of 
the two point correlation function 
from a pure power law shape are more serious 
in redshift-space than in real-space,
making it difficult to compare results obtained 
from the analyses of different galaxy samples.
To ensure a fair comparison, all best fit parameters listed in
columns 4 and 5 of
Table 1 
refer to ranges of separations where all measured
$\xi(s)$ are well approximated by a power law.
These ranges turned out to be very close to that of $[1-10]\hmpc$ considered in our analysis,
although some of the parameters in the Table have been obtained by
pushing the estimate of 
$\xi(s)$ down to scale as small 
as $0.1 \hmpc$ (the 2dFGRS case) 
or up to separations as large as $16.4 \hmpc$ (as in the case of the LCRS sample).

Our results are fully consistent with those obtained from the analysis of 
the PSC$z$ survey (Hawkins \etal 2001), 
that consists of $\sim 15,000$ {\it IRAS} galaxies
selected at $60 \mu m$, i.e. in a mid-infrared band similar to that of
ELAIS galaxies. This sample is, however, much more local
than ours and thus the agreement between the two results 
indicates that the clustering properties of mid-infrared selected objects
do not evolve significantly between $z=0$ and $z=0.3$.
This conclusion is
corroborated by the consistency found between the two measurements of $\xi(s)$ 
performed  in the {\it local} ($z<0.2$) and  {\it distant} ($0.2 \le z<0.5$) 
ELAIS-S1 subsamples.
Unfortunately, the large uncertainties in our estimate of $\xi(s)$,
that mainly result from the sparseness
of the ELAIS-S1 galaxy catalog,
do not allow to set strong constraints on the clustering evolution. 

The lack of significant 
evolution in the ELAIS galaxy clustering has already 
been noticed by  Gonzalez-Solares \etal (2004).
In their analysis, that consisted in 
deprojecting  the angular correlation function of ELAIS galaxies
via Limber equation, they have measured a real-space two point 
correlation function with a  
slope ($\gamma=2.04 \pm 0.18$) and a correlation length 
($r_0=4.3^{+0.4}_{-0.7} \hmpc$) that are fully consistent with those measured
in the PSC$z$ catalog ($\gamma=2.04 \pm 0.18$ and $r_0= \sim 3.7 \hmpc$,
Jing, B\"orner \& Suto (2002)).
It is worth stressing that difference between our result and that
of Gonzalez-Solares \etal (2004) originates from systematic redshift space distortions
that affect our analysis and 
result in a shallower slope and a larger correlation length of the
two point-correlation function.

Focusing on the mid-infrared objects is of considerable interest since they 
trace the underlying mass distribution in the local universe.
The consistency that we have found between the 
$\xi(s)$ in the real sample and the one measured in our mock ELAIS-S1 catalogs
in which galaxies are identified with the particles of the parent N-body simulation,
suggests that mid-infrared selected galaxies are indeed almost unbiased tracers of 
the underlying mass density field out to $z=0.5$.

It is interesting to compare our results with those obtained 
using optically-selected galaxy samples, some of which is 
listed in Table 1.
We confirm the well known fact  that infrared selected objects are significantly 
less clustered than optical galaxies, although the discrepancy depends  
on the depth of the sample and (mainly) on the selection band,
being larger for SDSS, $r^{\star}$ selected objects 
than for 2dF galaxies selected in the $b_j$ band.

Investigating how this discrepancy change with redshift, i.e. how
the relative clustering of different objects evolve with time, is of fundamental
importance since it will allow to cast light on the galaxy formation and evolution processes.
While this work constitutes an early
step in this direction, strong 
observational constraints for theories of galaxy evolution
will only be obtained by extending and improving 
samples like  ELAIS-S1, which implies
carrying out larger and deeper infrared redshift surveys
like that already planned in the SWIRE programme (Lonsdale \etal 2003).

\section*{Acknowledgements}

EB and FLF acknowledge ASI, INAF and MIUR (under grant
COFIN-03-02-23) for financial support.


\begin{thebibliography}{}
  \bibitem
    []{}Adelberger, K.L., 2003, ApJ, 584, 45
  \bibitem
    []{}Coil, A. \etal, 2004, ApJ, 609, 525
 
    \bibitem[]{}Cole, S., Hatton, S., Weinberg, D.H. \& Frenk, C.S., 1998, MNRAS, 300, 945


\bibitem[]{}Davis, M., \& Peebles, P.J.E., 1983, ApJ, 267, 465

\bibitem[]{}Davis, M. \etal, 2003, Proc. SPIE, 4834, 161 

\bibitem[]{}Efstathiou, G.,  1988, in Lawrence A., ed., Proc. 3rd {\it IRAS} Conf., 
Comets to Cosmology, Springer, New York, p. 312

\bibitem[]{}de Lapparent, V., Geller, M.J. \& Huchra, J.P., 1988, ApJ, 332, 44


\bibitem[]{}Gonzalez-Solares, A.E., \etal, 2004, MNRAS, 352, 44

\bibitem[]{}Gruppioni, C. \etal, 2002, MNRAS, 335, 831

\bibitem[]{}Gruppioni, C. \etal, 2005, ApJ, 618, L9

\bibitem[]{}Hamilton, A.J.S., 1993, ApJ,  417, 19

\bibitem[]{}Hawkins, E., Maddox, S., Branchini, E., Saunders, W., 2001, MNRAS, 
325, 589 

\bibitem[]{}Hawkins, E., \etal 2003, MNRAS,  346, 78

\bibitem[]{}Hermit, D.L. \etal  1996, MNRAS, 283, 709

\bibitem[]{}Hewett, P.C., 1982, MNRAS, 201, 867

\bibitem[]{}Jing, Y.P., B\"orner, G. \& Suto, Y., 2002, ApJ, 564, 15 

\bibitem[]{}Kerscher, M, Szapudi, I., \& Szalay, A.S., 2000,ApJL, 535, 13 

\bibitem[]{}Kessler, M.F. \etal 1996, A\&A, 315, L27

\bibitem[]{}La Franca, F. \etal 2004, AJ, 127, 3075

\bibitem[]{}Lahav, O. \etal 2002, MNRAS, 333, 961

\bibitem[]{}Landy, S.D. \& Szalay, A.S., 1993, ApJ, 412, 64

\bibitem[]{}Lari, C. \etal, 2001, MNRAS, 325, 1173



\bibitem[]{}Lonsdale, C.J. \etal, 2003, PASP, 115, 897

\bibitem[]{}Magdwick, D.S. \etal, 2003, MNRAS, 344, 847

\bibitem[]{}Matute, I. \etal 2002, MNRAS, 332, L11


\bibitem[]{}Norberg, P. \etal, 2002, MNRAS, 332, 827

\bibitem[]{}Oliver, S. \etal, 2000, MNRAS, 316, 749



\bibitem[]{}Pozzi, F. \etal, 2004, ApJ, 609, 122





\bibitem[]{}Rowan-Robinson, M. \etal 2004, MNRAS, 351, 1290

\bibitem[]{}Serjeant, S. \etal  2000, MNRAS, 316, 768

\bibitem[]{}Szapudi, I., Branchini .E., Frenk, C.S., Maddox, S., Saunders, W., 
2000, MNRAS, 318, L45

\bibitem[]{}Taylor A. N., Ballinger W.E., Heavens A.F. \& Tadros H.,
2002, MNRAS, 327, 689

\bibitem[]{}Tegmark M., Zaldarriaga M., \& Hamilton A.J., 2001, Phys. Rev. D,
67, 3007

\bibitem[]{}Tucker, D.L. \etal  1997, MNRAS, 285, L5


\bibitem[]{}Vanzella, E. \etal 2004, astro-ph/0406591

\bibitem[]{}Verde, L. \etal  2002, MNRAS, 335, 432

\bibitem[]{}Wirth, G.D. \etal 2004,  AJ, 127, 3121

\bibitem[]{}Zehavi, I. \etal 2002, ApJ, 571, 172


\end{thebibliography}
\end{document}